\documentclass[aps,prd,showpacs,nofootinbib,floats,floatfix,preprintnumbers,groupedaddress,twocolumn]{revtex4}

\usepackage{bm}
\usepackage{latexsym}
\usepackage{dcolumn}
\usepackage{amsmath,amsfonts,amssymb}
\usepackage{graphicx,epsfig}
\usepackage{amsmath}
\usepackage{fancyhdr}
\usepackage{hyperref}
\usepackage{graphicx,epstopdf}
\hypersetup{
	colorlinks   = true, 
	urlcolor     = blue, 
	linkcolor    = blue, 
	citecolor   = red 
}

{\scriptsize }

\begin{document}
\title{Symmetries near a generic charged null surface and associated algebra: An off-shell analysis}
\author{Mousumi Maitra\footnote {\color{blue} maitra.91@iitg.ernet.in}}
\author{Debaprasad Maity\footnote {\color{blue} debu@iitg.ernet.in}}
\author{Bibhas Ranjan Majhi\footnote {\color{blue} bibhas.majhi@iitg.ernet.in}}

\affiliation{Department of Physics, Indian Institute of Technology Guwahati, Guwahati 781039, Assam, India
}

\date{\today}

\begin{abstract}
To understand the underlying degrees of freedom, near horizon symmetry analysis of a black hole has gain significant interest in the recent past. In this paper we generalized those analysis first by taking into account a generic null surface carrying $U(1)$ electromagnetic charge. With the appropriate boundary conditions near the surface under study, we identified the symmetry algebra among the subset of diffeomporphism and gauge generators which preserve the metric of the null surface and the form of the gauge field configuration. With the knowledge of those symmetries we further derived the algebra among the associated charges considering  general Lanczos-Lovelock gravity theory and gauge theory. Importantly while computing the charges we not only considered general theory, but also used off-shell formalism which is believed to play crucial role in understanding quantum gravity. Both the nonextremal and extremal cases are addressed here.         

\end{abstract}

\maketitle

\section{Introduction}
Symmetries and the corresponding conserved charges play very important role in understanding the full dynamics of a theory. In this respect the theory gravity (with or without matter), has been one of the important areas of research for quite a long time mainly because of its non-linear nature. For a generic diffeomorphism invariant gravity theory, the Noether current and charge are shown to be very important in understanding the thermodynamic properties of black holes \cite{Iyer:1994ys}.  In this context, one of the important results is the commutator algebra among the charges associated with the asymptotic symmetries of a spacetime under study due to diffeomorphism. In general those lead to Virasoro algebra with a central charge \cite{Brown:1986nw}. This central charge is found to be intimately connected with the entropy of black holes through well known Cardy formula \cite{Strominger:1997eq}. 
Therefore, understanding the symmetries, specifically under the spacetime dependent transformation, has been central part in gravity for a long time (see \cite{Carlip:1999cy}--\cite{Bhattacharya:2018epn} for some later progress in this direction). 

In these endeavors, one recent study which gains significant attention is the exploration of symmetries near the asymptotic infinity and near the horizon. Asymptotic symmetries are such that it preserves the form the metric near the asymptotic boundary or  near the horizon. This analysis has been inspired by the old observations made by Bondi, Metzner, and Sachs \cite{Bondi:1962px}--\cite{Newman:1966ub}, in which the central question was to the understand the gravitational scattering phenomena and its effect near the asymptotic flat boundary. The symmetry which born out of their analysis is well known  BMS symmetry. This forms an infinite dimensional group which is a semidirect product of usual Poincare symmetry and the infinite dimensional super-translation symmetry transformation. This idea has been extended to different situations -- either for gauge fields \cite{Kapec:2015ena}--\cite{Campiglia:2017mua} or for gravity \cite{Barnich:2010eb}--\cite{Cai:2016idg} (for a review, see \cite{Strominger:2017zoo}). Therefore, the simplest idea would be to extend those BMS type analysis near the null horizon surface of black holes. The general strategy is to impose the conditions on the variation of the metric coefficients, expressed in Bondi-Sachs (or Gaussian null) coordinates, such that under diffeomorphism the near horizon structure does not change. Calculations have been done, till now, for a static or stationary horizon (see \cite{Donnay:2015abr, Eling:2016xlx, Akhmedov:2017ftb} for different cases) which are solutions of Einstein's equations of motion.

In this paper our aim is two fold. Firstly we study asymptotic symmetries and the associated algebra which keep a generic charged null surface invariant. In this paper we study  generic nonextremal and extremal charged null surface separately. To keep our study even more general while calculating various charges and their algebra, we also consider Lanczos-Lovelock (LL) theory of gravity in presence of arbitrary $U(1)$ gauge field. 
This will help us understanding the properties of charges while incorporating the interactions among the fields. Second, we will do our whole analysis {\it off-shell}. By this we mean that neither the Einstein's equations of motion nor the gauge field equation of motion will be used in our final results. Let us point out that this has not been looked at earlier. Now the questions are: Why are we interested in this and what will be achieved?

It has been observed that not only black hole horizon has thermodynamics interpretation, any generic null surface in gravity theory has this properties \cite{Parattu:2013gwa}. The idea stems from the equivalence principle-locally an accelerated frame known as Rindler frame can mimic gravity and hence it  can be a good candidate to explore various properties of gravity. Therefore, an accelerated observer in flat spacetime background is equivalent to a static observer in curved spacetime. This stimulates to think the gravity as an ``emergent phenomenon'' \cite{Padmanabhan:2009vy}. Nevertheless from the above discussions one would tend to think that understanding the  behavior of a generic null surface not only can provide the desired results of on-shell properties of the theory under study but also can shed light on the off-shell behavior which naturally appears in  quantum theory. 

Keeping this in mind, we  choose a generic null surface in the presence of gauge fields for our discussion. Imposing the relevant fall-off conditions for the metric coefficients and gauge fields, which asymptotically preserves the null structure, we find the associated diffeomorphism and gauge symmetry transformations. Then the algebra of the corresponding Fourier modes are obtained. The charges for the diffeomorphism and gauge symmetries are computed off-shell. We also computed the associated symmetry algebra for transformation parameters which are arbitrary function of null surface coordinates. It is quite evident that our analysis is completely off-shell not only by the choice of metric, but also by the derivation of charges as nowhere the information of equations of motion has been used. Also our results are valid for any order LL gravity in the presence of $U(1)$ gauge field. Here we consider both nonextremal and extremal situations. Hence we demand that our present analysis is much more general and must reflect the properties of a wide class of theories.

\section{Null surface: a brief introduction}
\label{null}
 In this section, we shall briefly discuss about the relevant properties of a generic null-hypersurface in an arbitrary spacetime dimension. For the description of the null surface we consider well known Gaussian null coordinate as $(u,r,x^A)$, with $A=2,3,\dots,d$ where $A$ corresponds to the angular coordinates. d is the number of spacetime dimension. The metric in this coordinate system is expressed as \cite{Hollands:2006rj,MORALES},
\begin{eqnarray}
ds^2=&&M(u,r,x^A) du^2 + 2 du dr + 2 h_A (u,r,x^A)dx^A du
\nonumber
\\
&&+ \mu_{AB} dx^A dx^B~,
\label{eqn1} 
\end{eqnarray}
where, we assume the null surface to be located at $r=0$.  The behavior of the metric components near the null surface are assumed as,
\begin{eqnarray}
M(u,r,x^A) = &&-2 \alpha(u,x^A) r + \mathcal{O}(r^2)~;
\nonumber
\\
h_A (u,r,x^A)= &&-r \beta_A(u,x^A) + \mathcal{O}(r^2)~;
\\
\mu_{AB}(u,r,x^A) = &&\mu_{AB}^{(0)}(u,x^A)+ 2 \mu_{AB}^{(1)}(u,x^A)r + \mathcal{O}(r^2)~. \nonumber
\label{eqn2}
\end{eqnarray}
At this point let us point out that the fall off condition for $g_{uu}$ has been chosen for nonextremal null surface. We will discuss extremal case separately at the end.
As has been emphasized in the Introduction, we will consider most generic null surface with all the metric components $M, h_A,$ and $\mu_{AB}$ being functions of all the spacetime coordinates ($u,r, x^A$). Moreover, we assume that the null surface is charged under the $U(1)$ gauge field ${A_{a}}$. Therefore, all the metric components in general will also be a function electric and magnetic charges. 
For a generic null surface we can define a null vector and its complimentary null vector $k^a = (1,0,0) $ and $l^a =(0,-1,0) $ respectively such that $g_{ab} l^a k^b=-1$ holds. For convenience we also mention here the covariant components of those two vectors as $k_a=(-2 r \alpha, 1, -r \beta_A)$, and $l_a=(-1,0,0)$. We also can see that the $r=0$ surface is a null $(d-2)$ dimensional sphere with an elementary surface area $d\Sigma_{ab}= - d^{(d-2)} x \sqrt{\mu}(k_a l_b-k_b l_a)$, where $\mu$ is the determinant of the induced metric on null surface. 

For generic charged null surface we also consider the following fall off conditions for the gauge field near the surface as,
\begin{equation} 
A_u = C^{(0)} +\mathcal{O}(r)   ; \,\,\ A_r = 0; \,\,\ A_B = C_{B}^{(0)}(u,x^A) +\mathcal{O}(r)~, 
\label{eqn3}
\end{equation}
where the time component of the gauge field assumes a constant value $C^{(0)}$ to the leading order in $r$. The time component of the gauge field is generally identified as a scalar potential. Therefore, it must be a constant on a particular surface. We also chose the following gauge condition, $A_r=0$. The choice of $A_B$ needs further explanation. However, for Kerr-Newmann black hole the condition appears to be true. As has been observed in \cite{Singh:2014paa,Kim:2014cja}, a static observe sitting outside the horizon will express the energy density of the electromagnetic field  as $U = T_{uu} (1/2 r \alpha)$, where $T_{uu}$ is the outgoing null-null component of energy-momentum tensor.
This component must be divergent as one approaches towards the null horizon $r=0$. Therefore it is sufficient to consider $T_{uu}$ to be finite.
 Considering the explicit expression for the electromagnetic energy momentum tensor one can immediately show that to the leading order in $r$, $ A_B \approx C_B^{(0)}(u,x^A)$. The detailed discussion on this issue is given in Appendix \ref{ap1}.

As emphasized before we study the near horizon symmetries of a generic charged null surface. Therefore, it generalizes a special class of null metric that can be obtained from the near horizon expansion of Kerr-Newmann black hole which is stationary (the procedure can be followed from \cite{Booth:2012xm}). Here we are considering the case with all the metric coefficients to be depending on all spacetime coordinates. Moreover, the null metric eq.(\ref{eqn1}) is not a solution of Einstein's equations of motion. Therefore our analysis will be much more general and covers a wide class of spacetimes. In that respect the present one differs from the analysis given in \cite{Donnay:2015abr,Akhmedov:2017ftb}.  

\section{Symmetries near the null surface}
\label{symm}
In order the understand the symmetry properties near a surface, the general approach is to define the appropriate fall-off conditions for the metric coefficients and the gauge field components. The appropriate fall-off conditions are such that it keeps all the gauge choices intact and remaining components of the metric and the gauge field assume the same form near the null surface $r=0$ after the symmetry transformations. For the present purpose we will simultaneously consider the symmetries under diffeomorphism and $U(1)$ gauge transformations. After the transformation we solve for the aforementioned boundary conditions and identify the appropriate generators and their algebraic properties. 

Let us first concentrate on the boundary condition of metric coefficients. These boundary conditions can be divided into two categories. One category is related to the gauge fixing conditions which we call ``strong'' ones,
\begin{equation}
\pounds_\zeta  g_{rr}= 0;\,\,\ \pounds_\zeta  g_{ur}=0;\,\,\,\ \pounds_\zeta  g_{Ar}=0~,
\label{eqn4}
\end{equation}
while remaining conditions are the ``weak'' ones, and those are
\begin{equation}
\pounds_\zeta  g_{uu} \approx \mathcal{O}(r);\,\,\ \pounds_\zeta  g_{uA} \approx \mathcal{O}(r);\,\,\  \pounds_\zeta  g_{AB} \approx \mathcal{O}(1)~.
\label{eqn5}
\end{equation}
In the above $\pounds_\zeta$ denotes the Lie derivative along the vector $\zeta^a$ corresponding to the diffeomorphism $x^a\rightarrow x^a+\zeta^a$.
These strong conditions says that the metric components which are zero or constant, must remain unaltered after the diffeomorphism and the weak conditions come from the leading order behavior of the metric coefficients $\alpha$, $h_A$, $\mu_{AB}$. 
Now as emphasized earlier, we also need to consider the behavior of the gauge field. The combined symmetry transformation of gauge and diffeomorphism will lead to the following transformation for the gauge field $A_{\mu}$ and satisfy,
\begin{equation}
\delta A_r = \pounds_{\zeta} A_r + \partial_r \epsilon = 0~, 
\label{eqn6}
\end{equation}
while the other components must satisfy,
\begin{equation}
\pounds_{\zeta} A_u + \partial_u \epsilon \approx \mathcal{O}(1);\,\,\ \pounds_{\zeta} A_B + \partial_B \epsilon \approx \mathcal{O}(1)~. 
\label{eqn7}
\end{equation}
In the above $\epsilon$ is the $U(1)$ gauge transformation parameter.  
Our aim now is to find out the diffeomorphism vector $\zeta^a$ and the gauge parameter $\epsilon$ which satisfy the above imposed conditions. This can be done in the following way. First we solve the strong conditions (\ref{eqn4}) to find different components of $\zeta^a$ and then we impose the weak conditions (\ref{eqn5}) on aforementioned solutions. From Eq.(\ref{eqn4}) we find 
\begin{eqnarray}
\zeta^u &=& F(u,x^A)~;\nonumber\\
\zeta^r &=& T(u,x^A)- r \partial_u F- \partial_B F \int r\beta^B dr~;\nonumber\\ 
\zeta^A &=& - \partial_B F \int \mu^{AB} dr + R^A(u,x^A)~,
\label{eqn8}
\end{eqnarray}
where $F$, $T$ and $R^A$ are the integration constants which are unknown at this moment (for a schematic derivation, please see Appendix \ref{ap2}).
The weak conditions for the metric component $g_{uu}$, $g_{uA}$, Eq.(\ref{eqn5})
give us the following constraint on the diffeomorphism parameters derived in Eq. (\ref{eqn8}):
\begin{eqnarray}
&&\partial_u T - \alpha(u,x^A) T= 0~;
\\
&&\partial_A T-T(u,x^A) \beta_A(u,x^A) 
+ \mu_{AB}^{(0)} \partial_u  R^B  =0~. \nonumber
\label{eqn10} 
\end{eqnarray}
The fall-of condition on $g_{AB}$ does not give any new condition (see Appendix \ref{ap2}).
Our goal is to understand the symmetry properties of the null surface located at $r=0$. Therefore, any transformation which changes the position of the null surface should vanish as we approach toward the null surface. This implies the vanishing $T(u,x^A)$ from  Eq.(\ref{eqn8}). Therefore, with the condition $T=0$, we just mentioned, all the constraint Eqs.(\ref{eqn10}) will automatically be satisfied provided rotation parameter $R^A$ being independent of $u$.
In a similar manner, the Eq.(\ref{eqn6}) solves for gauge parameter $\epsilon$. To find $\epsilon$ one has to use the components of $\zeta^a$ which have been found out by the conditions on $g_{ab}$.
Finally one finds the relevant parameters as
\begin{eqnarray}
&&\zeta^u = F(u,x^A)~;
\nonumber\\
&&\zeta^r = - r \partial_u F- \partial_B F \int r\beta^B dr~;
\nonumber\\ 
&&\zeta^A = - \partial_B F \int \mu^{AB} dr + R^A(x^A)~;
\nonumber\\
&&\epsilon = E(u,x^A) + \int_{r}^{} dr [A_B (\partial_C F) \mu^{BC}]~.
\label{eqn11}
\end{eqnarray}
Here $E$ is an another integration constant.
The above expressions are the symmetry parameters which keep the null surface structure invariant near $r=0$.
\section{Algebra of the symmetry parameters}
\label{algebra}
We are now interested to explore the algebra of the Fourier modes of the symmetry parameters very near to the null surface. In this case the nonvanishing parameters of our importance are
\begin{eqnarray}
&&\zeta^u = F(u,x^A); \,\,\ \zeta^{A_q}= R^{A_q}(x^{A_q})~;
\nonumber
\\
&&\epsilon = E(u,x^A)~.
\label{eqn12}
\end{eqnarray}
For our subsequent computation, we assume that $R^{A_q}$ are function of associated angular coordinate $x^{A_q}$. For example $\zeta^ \theta = R^\theta(\theta), \zeta^ \phi = R^ \phi(\phi)$, etc. $x^{A_q}$ is the $q^{th}$ angular coordinate. 
The Fourier modes of $F$, $R^A$, $E$ are expressed as,
\begin{eqnarray}
&&\zeta_{m,n}^u = F_{(m,n)} = \frac{1}{a} e^{i(m a u+ \sum_{A_q} n \ x^{A_q})}~;
\nonumber\\
&&\zeta_{k}^{A_q} = R_k ^{A_q}(x^{A_q}) = e^{i k \ x^{A_q}}~;
\nonumber\\
&&E_{(j,l)} = e^{i(j a u + \sum_{A_q} l \ x^{A_q})}~,
\label{eqn13}
\end{eqnarray}
where,$ m,n,k,j$ and $l$ are both positive and negative integers. $a$ is a constant having dimension of inverse  length. Hence the periodicity of the coordinate $u$ is taken to be $2 \pi / a$. The associated symmetry algebra of the aforementioned Fourier modes will be coming from the algebra satisfied by the various components of diffeomorphism vector $\xi^a$ and the gauge parameter $\epsilon$. 
The modified Lie bracket among those parameters are given by \cite{Barnich:2010eb},
\begin{equation}
[\zeta_1, \zeta_2]_M = [\zeta_1, \zeta_2] -  \delta^g_{\zeta_1} \zeta_2 - \delta^g_{\zeta_2}  \zeta_1~,
\label{eqn14}
\end{equation}
where,
\begin{equation}
[\zeta_1, \zeta_2]^x= \zeta_1^a \partial_a \zeta_2^x -\zeta_2^a \partial_a \zeta_1^x~.
\label{eqn15}
\end{equation}
 Here $\delta^g_{\zeta_1} \zeta_2$ denotes the variation in $\zeta_2$ under the variation of the metric induced by $\zeta_1$; i.e. due to $\delta^g_{\zeta_1} g_{ab}=\pounds_{\zeta_1}  g_{ab}$. However, from our analysis the components of the diffeomorphism vector $\zeta^a$ turned out to be independent of metric component as one approaches towards the null surface. So we can use only Lie bracket shown in Eq.(\ref{eqn15}), instead of the modified version of it.
Hence, the required symmetry algebra are constructed as follows,
\begin{eqnarray}
i \ \big[R_k^{A_q}, F_{m,n}\big] &=& -n \  F_{(m,n+k)} \nonumber\\
i \ \big[R_m^{A_q},R_n^{A_q'}\big] &=& (m-n) \  R^{A_q}_{(m+n)} \ \ \delta_{q,q'}\nonumber\\
i \ \big[F_{m,n},F_{p,q}\big] &=& (m-p) \  F_{(m+p,n+q)}\ \label{eqn16}\\
i \ \big[F_{m,n},E_{j,l}\big] &=& - j \  E_{(m+j,n+l)} \nonumber\\
i \ \big[R_m^{A_q},E_{j,l}\big] &=& - l \  E_{(j,m+l)} \nonumber\\
\big[E_{j,l},E_{m,n}\big] &=& 0~. \nonumber
\end{eqnarray}
Here it is clear that superroration vectors $(R_n^{A_q})$ are commutative for different angular variables. This is in sharp contrast with the usual rotation algebra. Moreover supertranslation vector $F_{m,n}$ is noncommutative with itself. This happens because of supertranslation generator $F$ being a function of both space and time coordinate. In this sense, our analysis is the generalization of studies reported in \cite{Donnay:2015abr}. In the subsequent section we will calculate various charges and their associated algebra corresponding to the diffeomorphism and $U(1)$ gauge symmetry transformations which keep our generic charged null surface invariant. 

\section{Charge and its algebra: an off-shell analysis}
\label{charge}
In this section our aim is to find the algebra among the Fourier modes of the charges corresponding to the aforementioned diffeomorphism vector and gauge parameter. For both the cases we shall first identify the most general expression for the Noether charges corresponding to general two derivative LL gravity in presence of matter and general $U(1)$ gauge invariant theories. The general action for gravity and minimally coupled $U(1)$ gauge invariant theory is taken to be, 
  \begin{equation}
{\cal L} = \int d^dx \sqrt{-g}\left(\frac{L(g_{ab},R^a_{~bcd}
)}{16 \pi G} + f(\mathcal{F}_{ab})\right)~,
\label{Am}
\end{equation}
where $L(g_{ab},R^a_{~bcd})$ corresponds to a general LL gravity theory. $f(\mathcal{F}_{ab})$ is a generic scalar function in terms of $U(1)$ field strength tensor $\mathcal{F}_{ab}=\nabla_aA_b-\nabla_bA_a$. For instance, in case of $U(1)$ Yang-Mills theory, it is given by $f=(1/16\pi)\mathcal{F}_{ab}\mathcal{F}^{ab}$. 

In this case, the charge due to both the diffeomorphism and gauge symmetries is given by $Q_{tot}=Q[\xi]+Q[{\epsilon}]$, where first term is the contribution originating  from the gravity while the other one is the matter part. Let us first concentrate on the gravity part.
For LL gravity, one obtains \cite{Paddy, Majhi:2011ws}
\begin{equation}
Q[\zeta]=\frac{1}{2}\int_{\mathcal H}d\Sigma_{ab}J^{ab}~,
\label{QL}
\end{equation}
where 
\begin{equation}
J^{ab}=\frac{1}{8\pi G}P^{abcd}\nabla_c\zeta_d~,
\label{Jab}
\end{equation}
with $P_{abcd}=\partial L/\partial R^{abcd}$. For the sake of generality we considered the above charge to be {\it off-shell} in the sense that one does not need to use the Einstein's equation of motion to derive this. This is very important for our purpose as we emphasized earlier that the generic null metric Eq.(\ref{eqn1}) under our present study does not need to be a solution of Einstein's equations of motion. Hence the algebra will of off-shell in nature. 

Since our charged null surface is locate at $r=0$, the surface integral will survive only for $d\Sigma^{ur}$ component. Therefore, the expression of diffeomorphism charges comes out to be,
\begin{eqnarray}
Q[\zeta]=\frac{1}{8 \pi G} \int_{\mathcal H}d\Sigma_{ur} P^{urcd} \nabla_c\zeta_d~.
\label{QG}
\end{eqnarray}
Using the symmetric properties of $P^{abcd}$ the above expression can be expanded as
\begin{eqnarray}
&&Q[\zeta]=\frac{1}{8 \pi G}\int_{\mathcal H}d\Sigma_{ur} [P^{urur} (\partial_u \zeta_r -\partial_r \zeta_u)+ P^{uruA}\nonumber\\ && (\partial_u \zeta_A -\partial_A \zeta_u)+P^{urrA} (\partial_r \zeta_A -\partial_A \zeta_r)\nonumber\\ && +\frac{1}{2} P^{urAB} (\partial_A \zeta_B -\partial_B \zeta_A)]~.
\label{QP} 
\end{eqnarray}
In the next step, we lowered the last two indices of $P^{abcd}$ so that it gives a nonzero finite value near the null surface. Finally using the explicit expression for the symmetry transformation parameter Eq.(\ref{eqn11}) and then taking the limit $r=0$ we found,
\begin{eqnarray}
&&Q[F,R^{A_q}]= - \frac{1}{8 \pi G} \int_{\mathcal H} d^{(d-2)} x \sqrt{\mu} \Big[P^{ur}_{~~ru} (2 \alpha F +2\partial_u F
\nonumber
\\ 
&& +\beta_A R^A) -P^{ur}_{~~uB} \mu^{AB} \partial_A F + P^{ur}_{~~rB}  \mu^{BC}  (\partial_u \mu_{AC})  R^A + \nonumber
\\ 
&& P^{ur}_{~~EF}  \mu^{EC}  \mu^{FD}  \Big(\partial_C (\mu_{DA}  R^A) -\partial_D (\mu_{CA}  R^A)\Big)\Big]~. 
\label{Q}
\end{eqnarray}
In terms of Fourier modes of the symmetry parameters $(\zeta^a, \epsilon)$ as shown Eqs.(\ref{eqn13}), the associated charges can be written as ,
\begin{eqnarray}
&&Q[F_{m,n}] = - \frac{1}{8 \pi G} \int_{\mathcal H} d^{(d-2)} x  \sqrt{\mu}  [P^{ur}_{~~ru} (2 \alpha F_{m,n} + 
\nonumber
\\ && 2\partial_u F_{m,n}) -P^{ur}_{~uB}  \mu^{AB}  \partial_A F_{m,n}]~,
\label{QF}
\end{eqnarray}
and
\begin{eqnarray}
 &&Q[R_k^{A_q}] = - \frac{1}{8 \pi G} \int_{\mathcal H} d^{(d-2)} x  \sqrt{\mu} [P^{ur}_{~~ru}  \beta_A  R^{A_q}_k + P^{ur}_{~rB}  \mu^{BC} 
 \nonumber
 \\ 
 && (\partial_u \mu_{AC}) R^{A_q} + P^{ur}_{~~EF}  \mu^{EC} \mu^{FD} [\partial_C (\mu_{DA} R^{A_q}_k)
 \nonumber
 \\ 
 && - \partial_D (\mu_{CA}  R^{A_q}_k)]]~.
  \label{QR}
 \end{eqnarray}

In a similar manner we will calculate the charge associated with the $U(1)$ gauge transformation. As has already been pointed out, our goal is to compute the charge for a general nonlinear $U(1)$ invariant Lagrangian. One such well known theory is called Born-Infeld electrodynamics with $ f(\mathcal F_{ab}) = \lambda^2 (-1+\sqrt{1+ \mathcal F_{ab}\mathcal F^{ab}/(8 \pi \lambda^2)})$. Where $\lambda$ is the Born-Infeld parameter. Clearly for large $\lambda$ limit one gets back the usual $U(1)$ electromagnetic theory. Goal is to keep our discussions as general as possible, therefore, we will not consider any specific form of $f(\mathcal F_{ab})$. The Noether current due to gauge symmetry is given by
 \begin{eqnarray}
 J^a = \nabla_b(f^{ab}\epsilon)~,
 \label{Ja}
 \end{eqnarray}
 where $\epsilon$ is the gauge parameter. An {\it off-shell} derivation of this current is presented in Appendix \ref{gauge}. Using Stoke's theorem and considering only the null boundary locate at $r=0$, one obtains the associated charge on the null surface as
 \begin{eqnarray}
 Q[{\epsilon}] = \int_{\mathcal H}  d \Sigma_{ab}  f^{ab}  \epsilon
 \label{Qgauge}
 \end{eqnarray}
 where, $f^{ab} = \partial f(\mathcal F)/\partial \mathcal F_{ab}$. This charge is also defined off-shell as no condition of the equation of motion has been imposed in the derivation.
 The  Fourier modes of the $U(1)$ gauge charge turns out to be
 \begin{eqnarray}
 Q[E_{m,n}] = \int_{\mathcal H} 2 d^{(d-2)} x  \sqrt{\mu}   f^{ur}  E_{m,n}~.
 \label{QE}
 \end{eqnarray}
We have all the three different types of charges for a generic charged null surface. Out of those $(Q[F_{m,n}],Q[R^{A_q}_{k}])$ are identified as the super-translation and super-rotation charges respectively. The associated symmetry transformations are responsible for connecting the two distinct null surfaces with different values of their characteristic data. Similarly we call $Q[E_{m,n}]$ as the super-gauge charge. Hence, associated symmetry transformation will change the electromagnetic charge of the null surface. In the solution space of the full Einstein equation, such as black hole spacetime, those conserved charges indicate the existence of soft hair near the horizon of the black hole. This has a potential to solve the so called information loss paradox of black holes (for recent discussion see \cite{Hawking:2016msc,Averin:2016hhm}).

Using the fundamental Lie bracket among the gauge parameters $[Q[\zeta_m], Q[\zeta_n]] = \pounds_{\zeta_m} Q[\zeta_n]$, the Lie bracket algebra among the various charges can be expressed as,
  \begin{eqnarray}
  i\big[Q[R_k^{A_q}], Q[F_{m,n}]\big] &=& -n \  Q[F_{m,n+k}]
  \nonumber\\
  i\big[Q[R_m^{A_q}],Q[R_n^{A_q'}]\big] &=& (m-n)  \delta_{q,q'} \ Q[R^{A_q}_{(m+n)}] 
  \nonumber\\
  i\big[Q[F_{m,n}],Q[F_{p,q}]\big] &=& (m-p)  Q[F_{m+p,n+q}]
  \nonumber\\
  i\big[Q[F_{m,n}],Q[E_{j,l}]\big] &=& - j  Q[E_{(m+j,n+l)}] 
  \nonumber\\
  i\big[Q[R_m^{A_q}],Q[E_{j,l}]\big] &=& - l  Q[E_{(j,m+l)}] 
  \nonumber\\
  \big[Q[E_{j,l}],Q[E_{m,n}]\big] &=& 0~.
  \label{B1}
       \end{eqnarray} 
 It is clear from the above Eqs.(\ref{B1}) that the symmetry bracket among the charges are isomorphic to that among diffeomorphism vectors. Here the gauge symmetry and the diffeomorphism symmetry together form a closed algebra is in sharp contrast with the usual transformation. The implication of this could be interesting to explore further.

At this point let us again emphasize the fact that our analysis does not depend upon the equation of motion of the fields under consideration. We started with a general charged null surface which is not a solution of Einstein's equation of motion. After this we follow the usual asymptotic symmetry analysis with a physically motivated fall of conditions of all the field under study near the surface. Associated with those symmetries we constructed conserved charges without taking into account the equation motion. Therefore, our  off-shell approach not only helps us to understand the symmetry properties of a generic null surface but also  be applies to the on-shell solution. therefore, it is much more general than the earlier analysis \cite{Barnich:2010eb, Donnay:2015abr, Barnich:2015jua, Barnich:2013sxa, Cai:2016idg}. 

Before we complete our analysis, we show that the same algebra can also be obtained from the Noether charge corresponding to the surface term of the gravitational action. For simplicity, we will only consider the usual Einstein-Hilbert action and its associated boundary term called  Gibbons-Hawking-York (GHY) surface term. The idea is the following. It is well known that the GHY term itself, calculated on the horizon, leads to horizon entropy. Moreover, its Noether charge plays the same role (see Sec. $2$ of \cite{Majhi:2015pra} for a detail discussion). The reason behind this is that both the terms will coincide on the null surface. Since we did not find such discussion in the literature, in Appendix \ref{surface} we show this similarity explicitly for a static spacetime.

The conserved Noether current for GHY term is given by \cite{Majhi:2012tf},
\begin{equation}
J^{a}[\zeta] =\nabla_b \ J^{ab}[\zeta]= \frac{1}{8\pi G} \nabla_b \ (K\zeta^a N^b - K \zeta^b N^a)~,
\label{KJ}
\end{equation}
where $N^a$ is the unit normal to the boundary $\partial V$ of a region of spacetime $V$. $K = -\nabla_a
N^a$ is the trace of the extrinsic curvature of this boundary surface and $J_{ab}$ is the Noether potential associated with diffeomorphism symmetry of the theory. 
Now since both the Noether charges of the Einstein-Hilbert action and GHY term leads to entropy when calculated on the horizon, we expect that the GHY Noether charge also lead to same algebra (\ref{B1}) for the parameters (\ref{eqn11}) obtained here.
For the given null surface (\ref{eqn1}) only relevant surface element is $d\Sigma_{ur}=d^{(d-2)}x$. The unit spacelike normal vector $N^a$ on $r=constant$ surface comes out as $N_a=(0, (r^2 \beta_A \beta^A +2r \alpha)^{-1/2},0)$ and 
so the contravariant components are as follows:
\begin{eqnarray}
N^a =\Big((r^2 \beta_A \beta^A +2r \alpha)^{-1/2}, \sqrt{r^2 \beta_A \beta^A +2r \alpha}, 
\nonumber\\
(r \beta^A / \sqrt{r^2 \beta_A \beta^A +2r \alpha})\Big)~.
\end{eqnarray}
Therefore, the trace of the extrinsic is calculated to be
\begin{eqnarray}
&&K= -(1/\sqrt{\mu}) \ [\partial_a (\sqrt{\mu})N^a + (\sqrt{\mu}) \ \partial_a N^a]\nonumber\\
&&= [\frac{1}{2} \mu^{AB} [ (\partial_u \mu_{AB})(r^2 \beta_A \beta^A +2r \alpha)^{-1/2} + (\partial_r \mu_{AB})\nonumber
\\
&& \sqrt{r^2 \beta_A \beta^A +2r \alpha} + (\partial_C \mu_{AB}) (r \beta^C/ \sqrt{r^2 \beta_A \beta^A +2r \alpha}) ] \nonumber
 \\
&& - \partial_u (r^2 \beta_A \beta^A +2r \alpha)^{-1/2} -\partial_r [\sqrt{r^2 \beta_A \beta^A +2r \alpha}]\nonumber
 \\
&&  - \partial_A [r \beta^A/ \sqrt{r^2 \beta_A \beta^A +2r \alpha}]]~.
\label{K}
\end{eqnarray}
Substituting all the relevant quantities in the charge expression for the parameters (\ref{eqn12}) one obtains 
\begin{equation}
Q[F,R^{A_q}]= \frac{1}{16 \pi G} \int_{H} d^{(d-2)} x \sqrt{\mu} [2 \alpha F +\partial_u F +\beta_A R^{A_q}] \label{QEN}
\end{equation}
which is the similar expression obtained earlier from the usual Noether charge [see Eqn. (\ref{Q})] with the value of $P^{abcd}$ for GR). Therefore it is obvious that this will also lead to the algebra (\ref{B1}). This further indicates that the surface term of gravitational action carries the information of the bulk theory (for more to this direction, see \cite{Majhi:2015pra} and the references therein). 

\section{null surface: extremal case}
As has been mentioned before, in this section we study symmetry algebra for a generic charged extremal null surface which is define as a zero temperature limit of a nonextremal null surface considered so far. In this case also we will perform the off-shell symmetry analysis for a generic gravity and $U(1)$ gauge invariant theory. The neighborhood of an extremal null surface is parametrized by Gaussian null coordinate as:
\begin{eqnarray}
ds^2 =&-&r^2 \alpha(u,r,x^A) du^2 + 2 du dr \\
&-& 2 r \beta_A (u,r,x^A) du dx^A
+ \mu_{AB} (u,r,x^A)dx^A dx^B~,\nonumber
\label{extremal} 
\end{eqnarray}
The extremality condition which is equivalent to zero temperature limit, is manifested into the fall off condition of $g_{uu}\sim{\cal O}(r^2)$ as one approaches towards the surface at $r=0$. Considering the following scale transformation $ r'= \lambda r$ ~~and $ u'=u/ \lambda$, the metric near the extremal null surface is approached by taking $\lambda \rightarrow 0$ ;
\begin{eqnarray}
&&ds^2=-r^2 \alpha(u,x^A) du^2 + 2 du dr - 2 r \beta_A (u,x^A) du dx^A
\nonumber
\\
&&+ \mu_{AB}(u,x^A) dx^A dx^B~.
\label{extremal} 
\end{eqnarray}
As one can observe the behavior of the remaining metric components $\alpha, \beta_A,$ and $\mu_{AB}$ are same as non-extremal surface defined in Eq.(\ref{eqn2}). The extremality  condition on the metric does not have any effect on the gauge field $A_{\mu}$ configuration near the surface Eq.(\ref{eqn3}). Therefore, given the metric and the gauge field configurations in the extremal null background, we will carry out the same analysis as before with the following modified fall off conditions, 
\begin{equation}
\pounds_\zeta  g_{uu} \approx \mathcal{O}(r^2);\,\,\ \pounds_\zeta  g_{uA} \approx \mathcal{O}(r);\,\,\  \pounds_\zeta  g_{AB} \approx \mathcal{O}(1)~.
\label{ext}
\end{equation}
As emphasized, all the remaining conditions remain the same. Therefore, the diffeomorphism parameters derived in Eq.(\ref{eqn8}) have to satisfy the modified constraint relations as follows,  
\begin{eqnarray}
&&\partial_u T = 0~;
\nonumber\\
&& T(u,x^A) \alpha + \partial^2_u F + h_A \partial_u R^A =0~;
\label{constraint}\\
&&\partial_A T-T(u,x^A) \beta_A(u,x^A) 
+ \mu_{AB}^{(0)} \partial_u  R^B  =0~. \nonumber
\end{eqnarray}
As stated earlier, radial component of the diffeomorphism vector $T$ must be zero near the surface. Therefore, all the constraint Eqs.(\ref{constraint}) will automatically be satisfied if we consider, 
\begin{eqnarray}
\partial_u R^A(u,x^A)=0~~;~~\partial^2_u F(u, x^A)=0~;\label{F}.
\end{eqnarray}
Hence, $R^A$ will be independent of $u$ coordinate. The general solution of $F(u,x^A)$ can be written as,
\begin{eqnarray}
F(u,x^A) = M(x^A) u + N(x^A)~;\label{SF}
\end{eqnarray}
We have two independent arbitrary functions $(M(x^A),N(x^A))$, in the time diffeomoprphism symmetry. 
In order to have regularity of the metric at $r=0$, an observer very near the surface must periodically  identify $u \rightarrow u+ \beta$, such that $\beta$ plays the role of inverse temperature. The same periodic identification for the aforementioned time diffeomorphism parameter $F(u,x^A)=F(u+ \beta,x^A)$ will
automatically fix one of the arbitrary functions $M(x^A)$ to be zero. Therefore, $F$ becomes independent of $u$, which is in sharp contrast with the nonextremal case described earlier.
Therefore, for extremal null surface, the asymptotic symmetry generators are-
\begin{eqnarray}
&&\zeta^u = F(x^A)~;
\nonumber\\
&&\zeta^r = - r \partial_u F- \partial_B F \int r\beta^B dr~;
\nonumber\\ 
&&\zeta^A = - \partial_B F \int \mu^{AB} dr + R^A(x^A)~;
\nonumber\\
&&\epsilon = E(u,x^A) + \int_{r}^{} dr [A_B (\partial_C F) \mu^{BC}]~.
\label{Fnew}
\end{eqnarray}
Following the same procedure as has been discussed for nonextremal null surface, the associated symmetry algebra will take the following form, 
 \begin{eqnarray}
 i \ \big[R_k^{A_q}, F_n\big] &=& -n \  F_{(n+k)}\nonumber\\
 i \ \big[F_m,F_n\big] &=& 0\nonumber\\
 i \ \big[F_n,E_{j,l}\big] &=& - j \  E_{(j,n+l)} \label{bracket}
 \end{eqnarray}
Henceforth we observe that for the extremal null surface, the supertranlation vector field $F$ commutes with itself which was noncommutative for nonextremal case (\ref{eqn16}).
Similarly algebra between charges will change only for those associated with supertranslation charges $Q[F_n]$ as,
\begin{eqnarray}
i\big[Q[R_k^{A_q}], Q[F_n]\big] &=& -n \  Q[F_{(n+k)}]\nonumber\\
i\big[Q[F_m],Q[F_n]\big] &=& 0\nonumber\\
i\big[Q[F_n],Q[E_{j,l}]\big] &=& - j  Q[E_{(n+l)}]\label{Ealgebra} 
\end{eqnarray}
Here also brackets among charges are isomorphic to that among the vector fields. Other results from (\ref{B1})remain exactly same as before. It would very interesting to understand the physical interpretation of this difference in symmetry algebra for two different null surfaces. How the zero temperature limit plays the role in determining the symmetry could be an interesting point to study.

\section{Conclusions}

One of the main goals of our present analysis was to understand the symmetry properties of a generic null surface defined in gravity theory minimally coupled with the electromagnetic gauge theory. As we have emphasized throughout our analysis, we have done two important generalizations of the existing analysis. In one direction we have considered the most general $U(1)$ invariant electromagnetic theory minimally coupled with a gravity theory at any arbitrary order in LL gravity.
On the other hand, in our derivation of symmetry algebra among the charges we make use of the off-shell formalism, where we have not considered any equation of motions. 
Therefore, our study can automatically gives the near horizon symmetry of any black hole of the theory under consideration. The analysis has also been extended to extremal null surface. As has been pointed out in the recent papers  \cite{Hawking:2016msc,Averin:2016hhm}, in the black hole background those near horizon symmetries are spontaneously broken. Therefore, in quantum theory those symmetry breaking will lead to the associated Goldstone modes which will behave as soft hairs. This may play important role in solving the black hole information loss paradox.    

Nonetheless, we found the near horizon symmetries which can be categorized as supertranslation and superrotation acting on the null surface under study. As discussed those transformations asymptotically preserve the structure of the null structure in the presence of gauge charge. Finally the algebra of the corresponding charges on the null surface have been computed. 
Our algebra for the parameters and the charges are different from the earlier ones. This difference is due to the fact that in general the super translational parameter can be function of null coordinate ($u$) and we have considered this situation. Whereas, the earlier ones did not take into account this. Therefore, our analysis takes care of most general situation in all respect. Hope this will illuminate more this paradigm.

\vskip 3mm
\noindent
{\it Note added}. Recently, a paper \cite{Setare:2018ziu} has been appeared in arXiv which deals the near horizon symmetries and associated algebra of the charges for a Kerr-Newman black hole. Our analysis is much more general as we considered here the charge for the any order LL gavity theory in presence of gauge fields with in general represented by the action which is any function of field strength. Moreover, the present one analyzed for any generic null surface, not restricted to stationary horizon. In addition, it is an off-shell analysis in all respect. Therefore our paper presents results which is much more general by nature.    
 
\vskip 3mm
{\bf Acknowledgment:}
One of the authors (M.M.) thanks Krishnakanta Bhattacharya for several useful discussions during the execution of the project.


\appendix
\section{Why is the angular component of gauge field $A_B \approx O(1)$ near the null surface?}
\label{ap1}
Here we are considering the static observer very near to the null surface $r=0$. It has been pointed out in the main text that for such an observer the energy density $U=T_{ab}u^a u^b$ diverges, where $u^a$ is the four velocity. Now for the given metric, $u^a$ is given by $u^a =(1/2 r \alpha, 0, 0)$, as $u^a u_a =-1$. Therefore energy density turns out to be
\begin{eqnarray}
U = \frac{T_{uu}}{2 r \alpha}~.
     \end{eqnarray}
In order for $U$ to be divergent, $T_{uu}$ should be finite as we approach toward $r=0$. Hence $T_{uu}$ is $r$ independent.

Now near the null surface the $U(1)$ electromagnetic gauge field  energy-momentum tensor will take,
\begin{eqnarray}
 \lim_{r\rightarrow 0}T_{uu} &=& \lim_{r\rightarrow 0}\mathcal F_{u c} \ \mathcal F^ c_{~u} - \frac 14 g_{uu} \mathcal F_{cd}\mathcal F^{cd} \nonumber\\
 &=& \mu^{AB} (\partial_u A_A)(\partial_u A_B)~.
\end{eqnarray}
Therefore, in order for $T_{uu}$ to be finite, $A_B \approx O(1)$.

\section{\label{ap2}Derivation of diffeomorphism and gauge parameters}
\subsection{Diffeomorphism vectors (\ref{eqn8})}
The first equation of (\ref{eqn4}) implies that 
\begin{equation}
2 g_{ur} \ \partial_r \zeta^u =0~,
\end{equation}
which immediately implies the form of $\zeta^u$ given in (\ref{eqn8}). Using this in the last condition of (\ref{eqn4}) one finds
\begin{equation}
 \mu_{AB} \  \partial_r \zeta^B + \partial_A F =0~.
\end{equation}
Solution of which leads to $\zeta^A$. Finally, use of these components in the second condition of (\ref{eqn4}) yields  
\begin{equation}
\partial_r \zeta^r + r \beta^A \partial_AF  + \partial_u F = 0~,
\label{1}
\end{equation} 
whose solution is the radial component of $\zeta^a$.	
 
\subsection{Equations in (\ref{eqn10})}
 Putting the components of $\zeta^a$ from (\ref{eqn8}) in the first condition of (\ref{eqn5}), we get near null surface as,
\begin{eqnarray}
\pounds_\zeta  g_{uu} = \partial_u T(u,x^A) - \alpha(u,x^A) T(u,x^A) .
\label{falloff}
\end{eqnarray}
In the above expression we have written down the leading order term. Now given the fall off condition $\pounds_\zeta  g_{uu}={\cal O}(r)$ as shown in Eq.(\ref{falloff}), the right-hand side of the above equation must vanished as it is ${\cal O}(1)$ in $r$.
Similarly near $r=0$, variation of the other metric components read
\begin{align}
\pounds_\zeta  g_{uA} = \partial_A T(u,x^A)-T(u,x^A) \beta_A(u,x^A)+
\nonumber\\
 \mu_{AB}^{(0)} \partial_u  R^B (u,x^A)~,
\end{align}
which must vanish according to $\pounds_\zeta  g_{uA} \approx O(r)$. This yields the other equation in (\ref{eqn10}). 
With these one can verify that the remaining conditions of Eq.(\ref{eqn5}) are automatically satisfied as one approaches toward the null surface. The variation of $g_{AB}$
\begin{align}
\pounds_\zeta  g_{AB} = F  \partial_u \mu^{(0)}_{AB}+ T \mu^{(1)}_{AB} + R^E(u,x^E)]  \partial_E   \mu^{(0)}_{AB} +
\nonumber\\ 
\mu^{(0)}_{AD} \ \partial_B R^D(u,x^D)~,
\end{align}
does not give us any new constraints  as it is already matching with assumed fall off condition $\pounds_\zeta  g_{AB}=\mathcal{O}(1)$.

\subsection{Derivation of gauge parameter $\epsilon$}
Using the derived forms of $\zeta^a$, the condition (\ref{eqn6}) leads to
\begin{eqnarray}
 A_B [-\partial_C F \ \mu^{BC}]+ \partial_r \epsilon =0
\end{eqnarray} 
whose solution yields the expression of gauge parameter $\epsilon$ given in Eq. (\ref{eqn11}).

The other conditions do not give any new constraints as they are now automatically satisfied. For instance one obtains near the null surface
\begin{eqnarray}
\pounds_{\zeta} A_u + \partial_u \epsilon= C_0 \ \partial_u F + A_B  \partial_u R^A + \partial_u E(u,x^A)~,
\end{eqnarray}
and
\begin{eqnarray}
\pounds_{\zeta} A_B + \partial_u \epsilon= F  \partial_u A_B + R^C \partial_C A_B  + (\partial_B F)  C_0 + 
\nonumber\\ 
A_C  \partial_B R^C + \partial_B E(u,x^A)~,
\label{g}
\end{eqnarray} 
which are $\mathcal{O}(1)$.

\section{\label{gauge}An off-shell derivation of gauge current}
The variation of the matter action (\ref{Am}) for an arbitrary change in field $A_a\rightarrow A_a+\delta A_a$ is given by
\begin{equation}
\delta {\cal L}_{em}= \int d^dx\sqrt{-g}\Big[\frac{\partial f}{\partial A_a}\delta A_a+ \frac{\partial f}{\partial(\nabla_a A_b)}\delta(\nabla_a A_b)\Big]~.
\end{equation}
Now since $f$ is function of $F_{ab}$ only, the first term will vanish. Denoting $\partial f/\partial F_{ab}=f^{ab}$, we find
\begin{eqnarray}
 \delta {\cal L}_{em} &=& \int d^dx\sqrt{-g}f^{mn}\delta(\nabla_m A_n-\nabla_nA_m)
\nonumber
\\
&=& 2\int d^dx\sqrt{-g}f^{mn}\nabla_m \delta A_n~.
\end{eqnarray}
Now if this variation is due to the gauge transformation $A_a\rightarrow A_a+\nabla_a\epsilon$, then the above equation reduces to
\begin{equation}
\delta {\cal L}_{em} =  2\int d^dx\sqrt{-g}f^{mn}\nabla_m \nabla_n\epsilon~.
\end{equation}
Since $f^{mn}$ is an antisymmetric tensor, the above equation can be expressed as a total derivative without using the equation of motion. The steps are as follows:
\begin{eqnarray}
&&\delta {\cal L}_{em} = 2\int d^dx\sqrt{-g}\Big[\nabla_m(f^{mn}\nabla_n\epsilon) - (\nabla_mf^{mn})(\nabla_n\epsilon)\Big]
\nonumber
\\
&&=2\int d^dx\sqrt{-g}\Big[\nabla_m\Big\{\nabla_n(f^{mn}\epsilon)-\epsilon\nabla_nf^{mn}\Big\}
\nonumber
\\
&&- (\nabla_mf^{mn})(\nabla_n\epsilon)\Big]
\nonumber
\\
&&=2\int d^dx\sqrt{-g}\Big[\nabla_m\nabla_n(f^{mn}\epsilon)-(\nabla_m\epsilon)(\nabla_nf^{mn})
\nonumber
\\
&&-\epsilon\nabla_m\nabla_nf^{mn}- (\nabla_mf^{mn})(\nabla_n\epsilon)\Big].
\end{eqnarray}
The third term vanishes as $f^{mn}$ is antisymmetric tensor while the second and last terms cancel each other. Therefore we are left with
\begin{equation}
\delta {\cal L}_{em} =2\int d^dx\sqrt{-g}\nabla_m\nabla_n(f^{mn}\epsilon) = 2 \int d^dx\sqrt{-g}\nabla_m J^m
\end{equation}
Now since the action has this gauge symmetry, its variation must vanish and hence we identify the conserved current as given in eq. (\ref{Ja}).

\section{\label{surface}Surface terms and corresponding Noether charge are same on horizon}
The entropy of the horizon is identified as 
\begin{eqnarray}
S&=&\frac{2\pi}{\kappa}\frac{1}{2} \int d\Sigma_{ab}J^{ab} 
\nonumber
\\
&=& \frac{1}{16\pi G}\int dtd^{d-2}x\sqrt{\mu}(N_aT_b-N_bT_a)J^{ab}~,
\end{eqnarray}
where the periodicity of the Euclidean time has been adopted in the last step. For horizon $N^a$ is the spacelike unit normal; i.e. $N^2=+1$ 
 while $T^a$ is the unit timelike normal: $T^2=-1$. $\mu$ is the determinant of the horizon induced metric. Next substituting the value of $J^{ab}$ from (\ref{KJ}) and using the fact that $\zeta^a$ is the timelike Killing vector, we find
\begin{equation}
S=\frac{1}{8\pi G}\int d^{d-1}x \sqrt{\mu} K(-T_a\zeta^a)=\frac{1}{8\pi G}\int d^{d-1}x\sqrt{h}K~, 
\end{equation}
where $h$ is the determinant of the induced metric on the radial coordinate constant surface which is timelike. This clearly shows that the two quantities, one is the GHY term and other one is the corresponding Noether charge multiplied by the periodicity of the Euclidean time, are same on the horizon. This is why both of them give the same quantity -- entropy of the horizon. 


\end{document}